\documentclass{article}

\usepackage{graphicx}
\usepackage{tocloft}
\usepackage{url}
\usepackage[colorinlistoftodos]{todonotes}
\usepackage[T1]{fontenc}
\usepackage[utf8]{inputenc}
\usepackage{siunitx}
\usepackage{color, soul}
\usepackage{algorithm}
\usepackage{algpseudocode}

\title{3D scattering microphantom sample to assess quantitative accuracy in tomographic phase microscopy techniques}

\author{Wojciech Krauze,$^{1,*}$ \and Arkadiusz Kuś,$^{1}$ \and Michał Ziemczonok,$^{1}$ \and Max Haimowitz,$^{2}$ \and Shwetadwip Chowdhury$^{2}$ \and Małgorzata Kujawińska$^{1}$}

\date{$^{1}$Warsaw University of Technology, Institute of Micromechanics and Photonics, Boboli 8 street, Warsaw, Poland, 02-525\\
$^{2}$University of Texas at Austin, Department of Electrical Engineering, 2501 Speedway, Austin, TX 78712, USA}

\begin{document} 
\maketitle

\begin{abstract}
In this paper we present a method to robustly evaluate the quantitative accuracy of various tomographic phase microscopy (TPM) methods with a multiple scattering 3D-printed microphantom with known geometry and refractive index distribution. We demonstrate this method by fabricating a multiple-scattering phantom and comparing 3D refractive index results that are output from three TPM reconstruction methods operating with visible and near-infrared wavelengths. One of these methods assumes the sample to be weak-scattering, while the other two take multiple scattering into account. This study can be readily extended to more complex microphantoms fabricated to more closely capture scattering characteristics of real-world scattering objects, such as tissue.

\end{abstract}

\section{Introduction}
\label{sect:intro}

Tomographic phase microscopy (TPM) is a quantitative, label-free imaging method that utilizes optical projections through a semi-transparent sample along various illumination angles to reconstruct the sample's 3D refractive index (RI). This method has found several applications in biological imaging, where RI is directly related to the dry mass distribution at the cellular and subcellular levels. Refractive index and dry mass are known to be crucial factors in analyzing the current stage of cell cycle \cite{girshovitz_generalized_2012}, cell structure \cite{kim2014common,stanly2020quantitative}, photobiochemical effects on cells \cite{Baczewska:22}, influence of external factors on cellular parameters \cite{baczewska2021refractive, eder2021medical} and many others. However, to reconstruct 3D RI, traditional TPM methods utilize critical assumptions in their computational reconstruction methodologies that rely on the sample being \textit{weakly scattering} \cite{chen1998validity}. These assumptions limit samples to having thicknesses on the order of only tens of microns. Given the wealth of information provided by dry mass at the single-cell level, there is significant demand to analyze dry mass in large multicellular clusters, thick tissue slices, or whole microorganisms. For these types of samples, reconstruction frameworks that accommodate for \textit{multiple scattering} must be utilized. To this end, numerous TPM approaches have been proposed in recent years that introduce new frameworks to accommodate multiple scattering \cite{lee2022inverse, liu2017seagle, sun2018efficient, soubies2017efficient, chen2020multi, Chowdhury2019}. Notably, these approaches utilize nonlinear and nonconvex solvers to iteratively solve for a sample's 3D RI. Though these methods have demonstrated impressive results in reconstructing RI in multiple-scattering samples, their quantitative accuracy has not yet been robustly characterized experimentally, and the presented results usually do not allow comparison of different methods in order to select proper approach for a given scattering level in a sample. The general strategy to experimentally evaluate quantitative accuracy in TPM methods is to reconstruct 3D RI in samples with known RI distributions \cite{lim2015comparative,sung2011deterministic,lim2017beyond}. Unfortunately, existing microphantoms are typically either weak-scattering  (e.g. index-matched microspheres) or overly simplistic  (e.g. index-mismatched microspheres) compared to the types of heterogenously scattering multicellular samples that multiple-scattering methods are intended for. This is a critical limitation when characterizing TPM reconstruction methods that utilize nonconvex solvers, where iterative convergence depends on the complexity of the energy landscape and directly associates with a sample's 3D complexity \cite{hindi2004tutorial}. A multiple-scattering TPM method that outputs accurate 3D RI reconstructions of a weakly scattering or overly simplistic microphantom cannot be expected to output similarly accurate RI reconstructions for more complex multiple-scattering samples, where the probability of converging to local minima are drastically higher. To robustly characterize the quantitative accuracy of multiple-scattering TPM methods, it is imperative to use microphantoms with known 3D RI that mimic the structural complexity of the types of samples that the TPM methods are intended to image. \textit{To the best of our knowledge, these types of gold-standard multiple-scattering microphantoms do not exist.} \par 
In this work we present a 3D-printed microphantom with internal multiple-scattering RI distribution. To do so, we leverage recent developments in 3D printing via direct laser writing \cite{lamont2020direct, lafratta2017two, horng20213d, eifler2018calibration}. We specifically utilize a two-photon polymerization technique that enables 3D printing of microphantom samples with known geometry and calibrated RI. Additionally, we use these microphantom samples to quantitatively assess the RI reconstruction quality in three TPM reconstruction algorithms that have demonstrated impressive biological imaging results in prior works. 

\section{Methods}
\label{sec:methods}
We present below the 1) methodology and design with which we 3D-print multiple scattering microphantoms; 2) the optical design of the TPM imaging systems that we use to experimentally collect scattering electric-field measurements of the microphantom; and 3) short theoretical descriptions of three tomographic algorithms that were used to reconstruct 3D RI from the measured data. 

\subsection{Design and manufacturing of the 3D scattering microphantom}
\label{sec:methods-3dphantom}
The phantom is fabricated using two-photon laser lithography, in which a focused laser beam is scanned within liquid resin. The resin within the laser's focal volume is locally polymerized. Adjusting the scanning trajectory and the exposure time of the laser beam enables simultaneous control over the 3D printed geometry (accuracy at the order of 100 nm) and RI (accuracy at the order of $5\cdot10^{-4}$, maximal $\Delta$RI = 0.03 within the structure) in three dimensions. We used Photonic Professional GT (Nanoscribe GmbH) equipped with a 1.3 NA 100x microscope objective and piezo scanning stage. The phantom is fabricated in the IP-Dip resin (Nanoscribe GmbH) on top of a \#1.5H coverslip (dip-in configuration \cite{Bunea2021_DiLL}). After fabrication the structure was developed in PGMEA (Propylene glycol monomethyl ether acetate; 12 min), followed by isopropyl alcohol (10 min) and then blow-dried. \par 

\begin{figure}
    \centering
    {\includegraphics[width=0.8\linewidth]{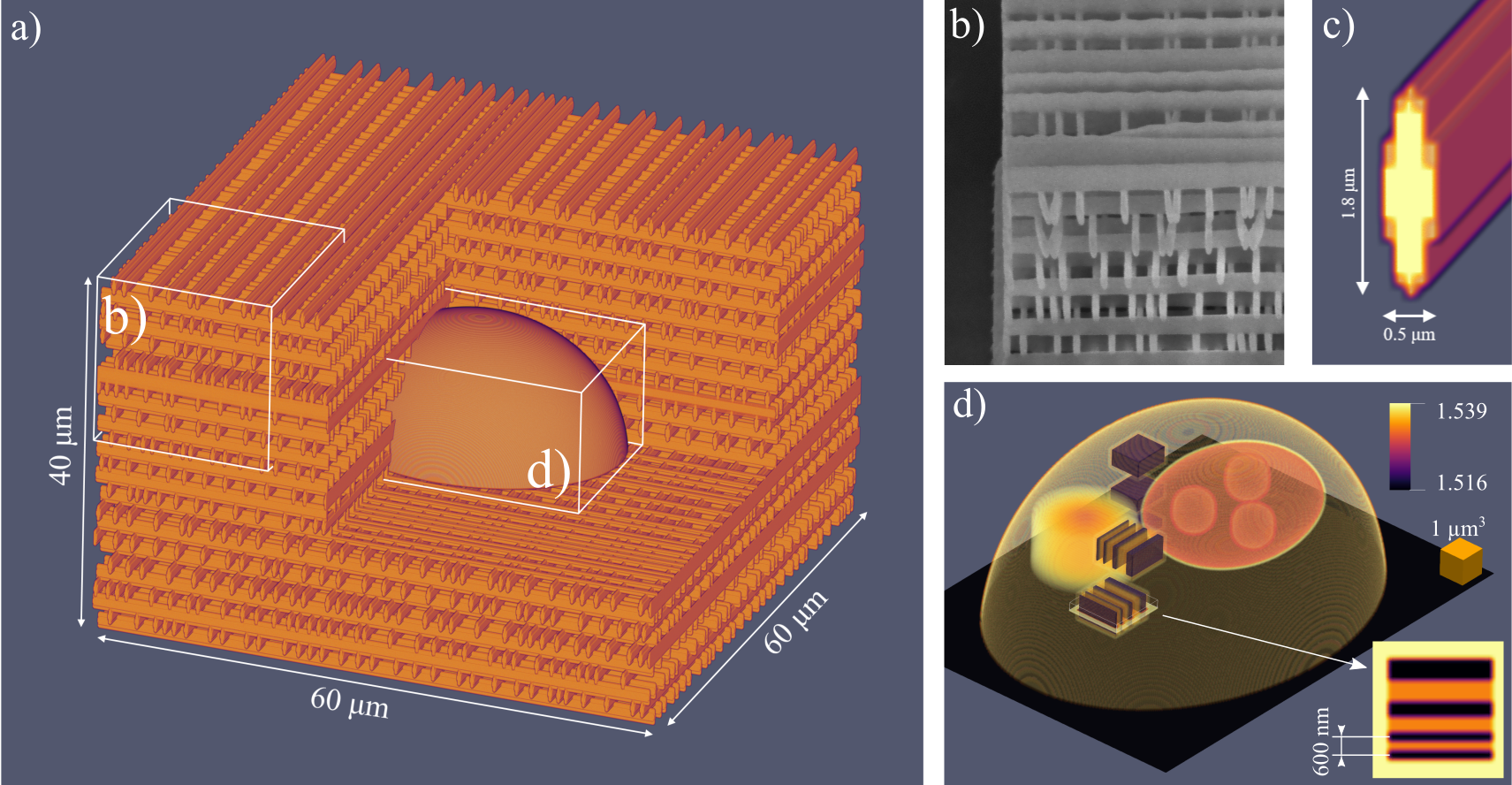}}
    \caption{a) Half-section view of the scattering microphantom design. b) SEM image of a scattering layer made out of quasi-randomly distributed rods. c) Individual rod that comprise a scattering layer. d) Visualization of the 3D RI distribution of the imaging target -- cell phantom. Subcellular features, such as resolution targets (shown in the inset) and cell nucleoli, are enclosed in truncated ellipsoid with the external dimensions of $\SI{30}{\um} \times \SI{25}{\um} \times \SI{12}{\um}$, which is then embedded in the center of the $\SI{60}{\um} \times \SI{60}{\um} \times \SI{40}{\um}$ scattering cube.}
    \label{fig:phantom_design}
\end{figure}

For the purposes of demonstration, the particular microphantom we 3D printed here consists of a cell-like target with internal test structures \cite{phantom_2019} embedded within a pseudo-random distribution of rods that switch their orientation across various layers (Fig. \ref{fig:phantom_design}a). The width and height of each rod is equal to \SI{0.5}{\micro \meter} and \SI{1.8}{\micro \meter} respectively. The lateral distance between the rods in each layer is randomized between \SI{0.7}{\micro \meter} to \SI{3}{\micro \meter} and the layers are stacked vertically every \SI{1.4}{\micro \meter}. The resulting structure is transparent (over 99\% transmittance for the extinction coefficient of \SI{0.1}{\milli\meter^{-1}} \cite{Schmid2019}) and multiple scattering \cite{scattering_2020}. The final scattering region is a $\SI{60}{\um} \times \SI{60}{\um} \times \SI{40}{\um}$ cube with a fill-factor of $\sim$25\% (fraction of volume occupied by the polymer).  \par 

The cell-like target embedded within the layers of randomly distributed rods mimics a single biological cell encased within a scattering cube. The cell-target comprises of substructures that enable assessment of quantitative accuracy in TPM imaging systems \cite{Ziemczonok2022_measurement}.  Methodology for fabrication and validation of these features can be found in our previous work \cite{phantom_2019}. The main features within the cell target include resolution test targets, nucleoli suspended in a nucleus, and a region of slow RI variation (see Fig. \ref{fig:phantom_design}d). Notably, the resolution test target comprises of lines with increasing spatial frequency \cite{Horstmeyer2016} up to 1667 lp/mm. By assessing the maximum spatial frequency of lines that can be distinguished within the cell test target, the imaging resolution of a TPM method of choice can be characterized. The pseudo-random distribution of rods that compose the scattering portion of the whole microphantom are suppressed within \SI{0.5}{\micro \meter} of the cell target and do not intersect with any of the test structures. To conduct our TPM imaging experiments, the microphantom was immersed in Zeiss Immersol 518F oil (RI$_{\SI{632}{\nano\meter}}$ = 1.5123), which provides similar RI contrast as in the case of cells immersed in culture medium. By using immersion oils with varying RI, it is possible to adjust the scattering properties of the microphantom post-fabrication. 

\subsection{Experimental setup}
\label{sec:methods-opticalsetup}
\begin{figure}[H]
    \centering
    \includegraphics[width=1.0\textwidth]{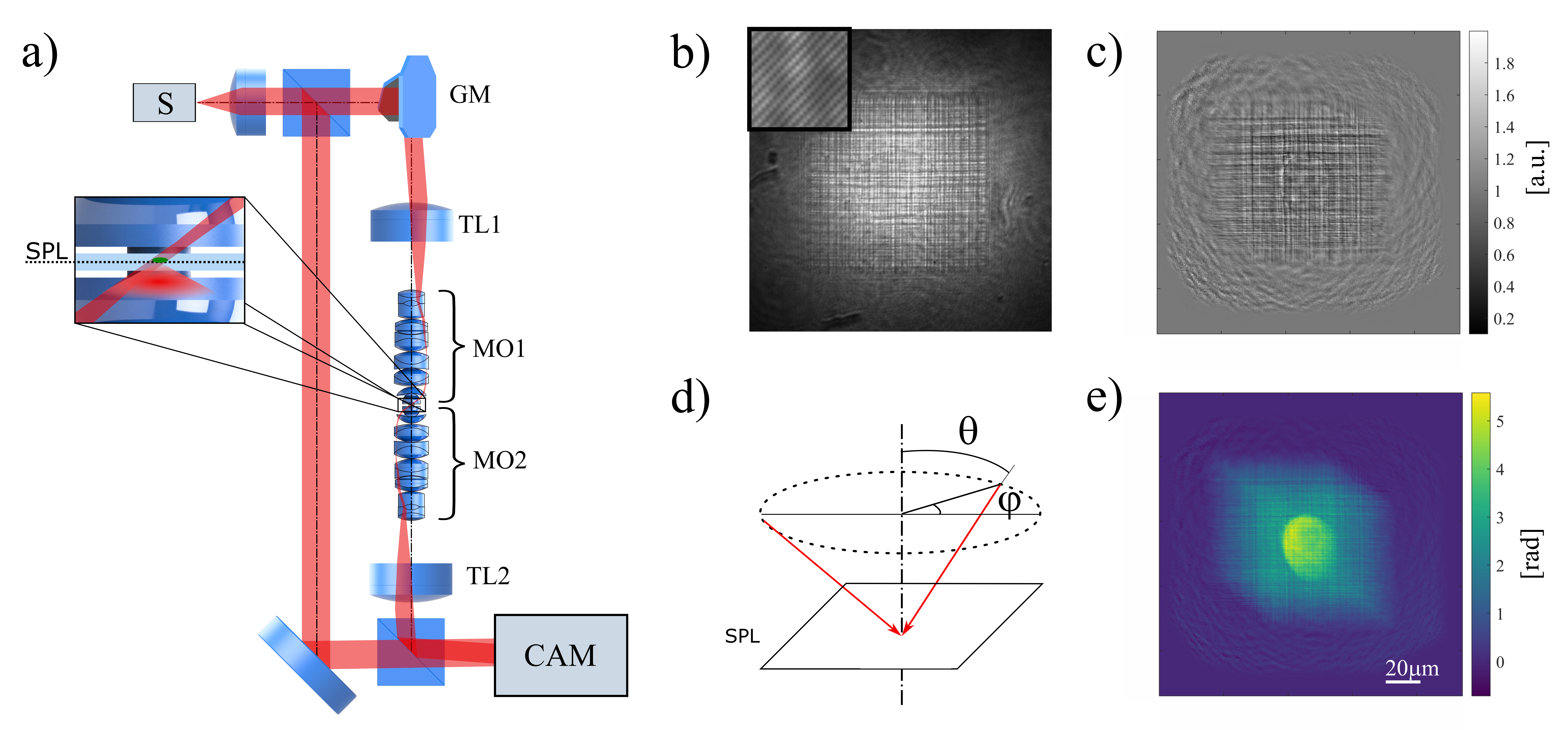}
    \caption{a) Mach-Zehnder-based TPM measurement system.  S - light source, GM - galvo system, TL1 and TL2 - tube lenses, MO1 and MO2 - microscope objective, SPL - sample plane, CAM - camera; b) hologram acquired at axial illumination of the sample; c) amplitude of a projection at $\varphi=304$\textdegree; d) circular scanning scenario used in the measurement. Rotation angle: $\varphi$, zenith angle: $\theta=47$; e) phase of the projection.}
    \label{fig:MeasSys}
  \end{figure}

In this work, an optical system, as shown in Fig. \ref{fig:MeasSys}a) was used in order to study the scattering phantom. The system is a Mach-Zehnder-based TPM microscope\cite{Kus2017}, working in a limited-angle configuration with stationary sample and illumination rotated with a galvo mirror (Thorlabs GVS212/M) \cite{Kus2019}. The research was performed with two wavelengths and thus there were two modified versions of the presented tomographic microscope. First version, $\text{TPM}_{633}$ works with wavelength $\lambda=$ \SI{633}{\nano \meter} and the second, $\text{TPM}_{835}$ with $\lambda=$ \SI{835}{\nano \meter}. The input beam (S in Fig.\ref{fig:MeasSys}a) is delivered with an optical fiber, collimated and then split into object and reference arm. In the $\text{TPM}_{633}$ system the light source was a volume Bragg grating laser (Necsel NovaTru Chroma 633 SLM), $\text{S}_{633}$, providing a single longitudinal mode and offering long coherence length. The $\text{TPM}_{835}$ system utilized a swept source (Superlum Broadsweeper BS-840-2-HP, $\Delta\lambda$= 800-870 nm), $\text{S}_{835}$ set at $\lambda=$ \SI{835}{\nano \meter}. Due to difference in coherence length, an additional delay module was placed in the reference beam for $\text{TPM}_{835}$ measurements. The beam-splitting cubes in this work were either coated for 400-700nm or 700-1100nm depending on the wavelength used. The focal length of the tube lens TL1 was $\text{EFL}_{633}\text{=}$ \SI{150}{\milli \meter} and $\text{EFL}_{835}\text{=}$ \SI{200}{\milli \meter} respectively. Both microscope objectives (MO1 and MO2) in Fig. \ref{fig:MeasSys} were 100x NA 1.3 Semi plan-apochromatic, infinity-corrected objectives. The second tube lens TL2 used was either $\text{EFL}_{633}\text{=}$ \SI{200}{\milli \meter} or $\text{EFL}_{835}\text{=}$ \SI{300}{\milli \meter}. This provided magnifications $M_{633}=-48.5$ and $M_{835}=-72.7$. 
The camera used in the system was a CMOS sensor in both cases, with 3.45$\mu m$ pixel size (JAI BM500GE) in case of $\text{CAM}_{633}$ and 5.5$\mu m$ pixel size (Basler acA2040-180km) in case of $\text{CAM}_{835}$.
The minimum magnification, which is imposed by pixel size and wavelength in order to assure correct hologram recording for each projection \cite{Sanchez-Ortiga2014} is $M_{633 min}=-44.2$ and $M_{835 min}=-53.5$, which is satisfied in both cases. A sample hologram is presented in Fig.\ref{fig:MeasSys}b).
Both systems were set to illuminate the sample with a circular scanning scenario (see Fig. \ref{fig:MeasSys}d ) at zenith angle $\theta=47$\textdegree and provided 180 projections spaced at $\varphi=2$\textdegree.
A sample of the phase and amplitude provided by the system at 835nm is presented in Fig.\ref{fig:MeasSys} c) and e).

\subsection{Reconstruction algorithms}
\label{sec:methods-algorithms}
We compare three different TPM reconstruction methods for computing 3D RI of the microphantom from the multi-angle scattering measurements captured as described above. We provide a short description of these methods below. Complete descriptions of these methods is given in respective references.

\subsubsection{Gerchberg-Papoulis with object support}
To provide a baseline standard to compare to TPM reconstruction algorithms utilizing multiple-scattering models, we first reconstruct the microphantom's 3D RI using a weak-scattering TPM method. We specifically use the Gerchberg-Papoulis algorithm enhanced with additional finite object support constraint (GPSC) \cite{krauze2020optical}. Complex-valued electric-fields measured by our TPM systems are used as inputs. The procedure is performed in two steps. First, an initial tomographic 3D RI distribution is reconstructed from the electric-field measurements with strong total-variation regularization \cite{strong2003edge}. This is performed through the Chambolle-Pock \cite{chambolle2011first} optimization method and implemented with ASTRA tomography toolbox \cite{van2015astra}. The result undergoes binarization and a finite object support is generated. Secondly, a classic Gerchberg-Papoulis algorithm is used which is an iterative version of Direct Inversion method (also known as the Wolf transform) \cite{wolf1969three}. This iterative procedure is based on the Fourier Diffraction Theorem \cite{Haeberle2010} and utilizes first-order scattering approximation. Here, the reconstruction and its Fourier transform are calculated alternately and constraints are applied: nonnegativity and finite object support in the signal domain, and replenishment of original projections in the frequency domain.

\subsubsection{Multi-slice beam-propagation with electric-field measurements}
We utilize the multi-slice beam-propagation method (MSBP) as our main method to model multiple-scattering \cite{van1981beam}, which has recently shown promising results for biological imaging \cite{Kamilov2015,kamilov2016optical}. In our first implementation of MSBP, we use the same exact electric-field dataset utilized by GPSC from above. An initial guess of the sample's 3D RI is selected to start off the iterative procedure. Afterwards, the MSBP method is used to simulate scattering measurements resulting from plane waves propagating through the sample's initial estimated 3D RI. The scattered fields resulting from this simulation are compared with those obtained experimentally with our TPM systems. The error computed between the simulated and experimental measurements is back-propagated through each layer of the 3D sample estimate to incrementally modify the RI value of each voxel. Continued iterations repeating these steps eventually result in the 3D sample estimate converging to a stable steady-state solution. We implemented the MSBP with electric-field measuemerements (MSBP-E) through the Learnining Tomography algorithm (LT) \cite{Kamilov2015}. The LT procedure is an iterative optimization algorithm with additional weak total-variation (TV) regularization applied in each iteration to ensure convergence. We found that the method works best when an initial-guess is chosen as a starting point for the iterative process. In this paper, we use the Direct Inversion method to provide the initial guess.

\subsubsection{Multi-slice beam-propagation with intensity-only measurements}
Recent works have demonstrated that the gradient-update step within the MSBP method can be reformulated to reconstruct 3D RI from only \textit{non-interferometric intensity measurements} \cite{Chowdhury2019}. The key advantages of this method include the use of a non-interferometric imaging system, which are resistant to mechanical instabilities that often limit long-term use of dual-arm interferometers without realignment. Furthermore, the light source can be partially coherent, to avoid coherent speckle artifacts in the measurements, while retaining sufficient coherence necessary for RI reconstruction. For the purposes of demonstrating 3D RI reconstruction using this intensity-only variant of MSBP (which we refer to here as MSBP-I), we simply use the amplitude component of the electric-field measurements used for the GPSC and MSBP-E reconstructions, described above.  Similarly to MSBP-E, total-variation regularization is applied at every iteration. The starting point for MSBP-I is a matrix of zeros.

\subsubsection{Stopping criterion}
All described TPM reconstruction methods are iterative procedures that use the same stopping criterion to automatically terminate the computations. This criterion is a modification of a method presented earlier \cite{Makowski:19}, and is described in Alg. \ref{alg:stop} below. The general intuition behind this procedure is to terminate the iterative computation process when the dynamics of the change between 3D sample estimates outputted from consecutive iterations drops below a certain saturation level $\epsilon$. In order to be less dependent on outliers, the median value of the dynamics from the last 10 iterations is calculated. The value for $\epsilon$ is chosen empirically for each algorithm. For GPSC $\epsilon=0.02$, for LT and MS $\epsilon=0.01$.

\begin{algorithm}
\caption{Stopping criterion}\label{alg:stop}
    \begin{algorithmic}
        \For{$i=1:\textit{NIter}$} \Comment{\textit{NIter} - max number of iterations}
        {
        \State Calculate reconstruction $x_i$ with a given method
        \If{$i > 2$}
            \State $d_i=||x_i - x_{i-1}||_2 / ||x_i||_2$ \Comment{$d_i$ - "distance" between \textit{i}-th and \textit{(i-1)}-th reconstruction}
        \EndIf
        \If{$i \geq 10$}
                \State $p_i = |d_i - d_{i-1}| / |\widehat{d(3:10)}| $ \Comment{$p_i$ - dynamics of "distance"}
        \EndIf  \Comment{$\widehat{(...)}$ - median value}
        \If{$\widehat{p(\text{end}-11:\text{end})}<\epsilon$}    \Comment{$\epsilon$ - saturation level}
            \State $i = \textit{NIter}$
        \EndIf
        }
        \EndFor
    \end{algorithmic}
\end{algorithm}

\section{Results}
\begin{figure}
\makebox[\textwidth][c]{\includegraphics[width=1.2\linewidth]{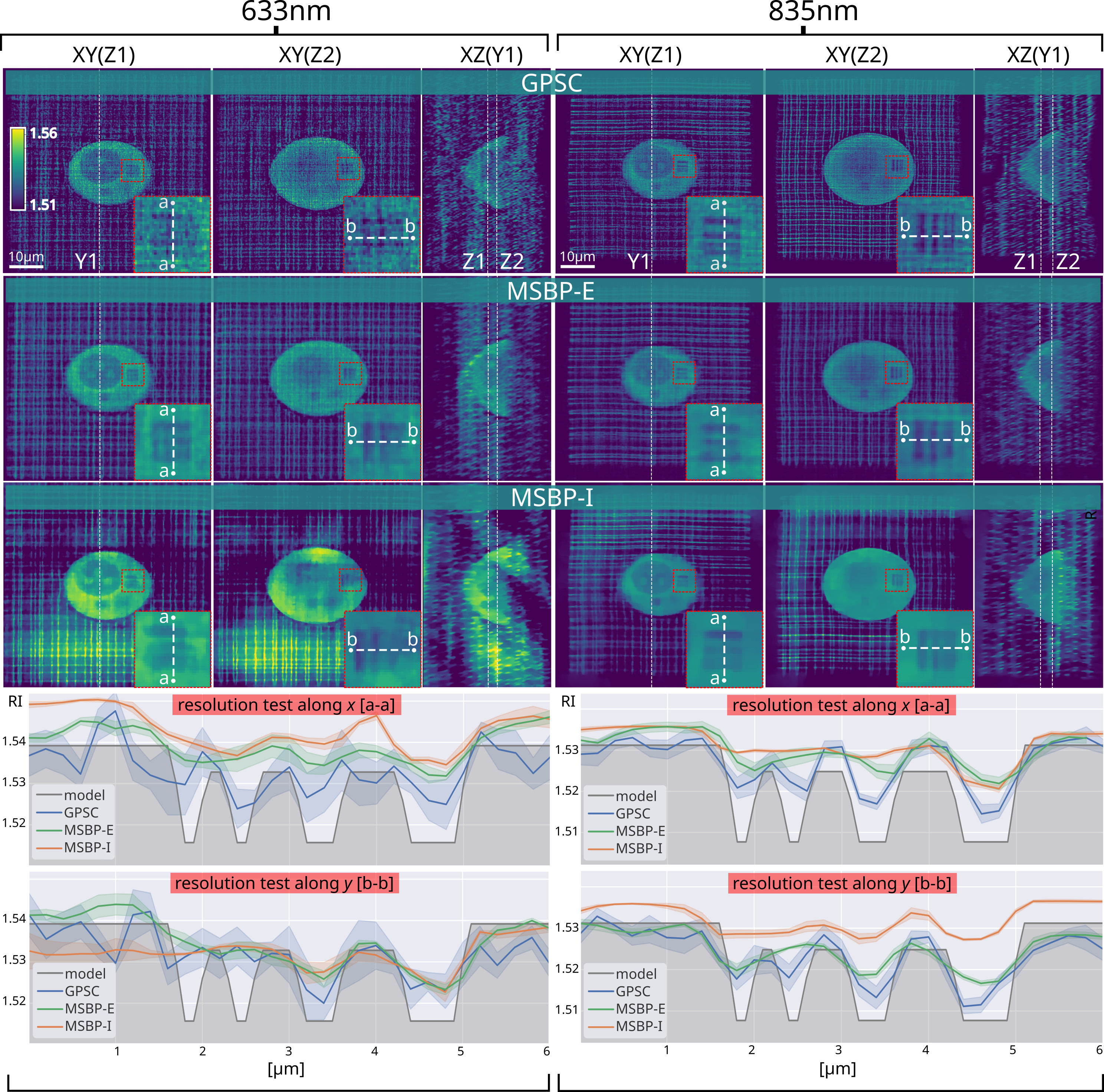}}%
\caption{Comparison of tomographic reconstructions of the microphantom measured with \SI{633}{\nano \meter} and \SI{835}{\nano \meter} wavelength and calculated with 3 algorithms. The shaded colored regions surrounding each of the 1D plots at the bottom represent the standard deviation.}
\label{fig:results-combined}
\end{figure}

Two sinograms were composed from the complex-valued scattered field measurements of the microphantom being illuminated at different angles, using both \SI{633}{\nano \meter} and \SI{835}{\nano \meter} wavelengths. For GPSC and MSBP-E, these sinograms were used directly. For MSBP-I, only the amplitude components of these sinograms were used. The sinograms and their corresponding 3D RI reconstructions \textbf{have been made open-source and are available in Dataset 1} \cite{zenodo}. Figure \ref{fig:results-combined} shows the reconstruction results for both wavelengths. We characterize lateral (\textit{x} and \textit{y}) resolution of the TPM reconstructions by visualizing the resolution test lines inside the microphantom. To quantitatively compare the three TPM reconstruction methods, horizontal and vertical cross-sectional plots across these resolution tests were generated by computing the average and standard-deviation of the pixel-values across rows or columns adjacent to the dashed white \textit{a-a} and \textit{b-b} lines, by $\pm 4$ pixels. These cross-sectional plots are shown below.

\section{Discussion}

The results show that changing the wavelength from visible into near-infrared impacts the applicability of TPM reconstruction methods \cite{helmchen2005deep, rubart2004two}. The magnified views of the resolution test region after 3D reconstructing the microphantom using the GPSC algorithm with \SI{633}{\nano \meter} wavelength light reveals significant grainy artifacts which occlude the line features within the microphantom's test region. We note that these artifacts are heavily decreased when reconstructing with \SI{835}{\nano \meter} wavelength light. This suggests that, at \SI{633}{\nano \meter} wavelength, the microphantom is too highly-scattering  for GPSC to be applicable. The decreased GPSC reconstruction artifacts for \SI{835}{\nano \meter} matches conventional knowledge that longer wavelengths of light are more resistant to scattering than shorter wavelengths. In order to confirm that this observation is due to reduced scattering and not different noise characteristics between the two light sources, we analyzed the standard deviation of phase-noise in an object-free region within the input sinograms. For \SI{633}{\nano \meter} wavelength light, a phase-noise standard deviation of $\sigma = 0.10$ radians was observed, while for \SI{835}{\nano \meter} wavelength light, a phase-noise standard deviation of $\sigma = 0.08$ radians was observed. Given such a small variation between these noise characteristics, we conclude that the major factor in the GPSC reconstruction quality that affects its ability to visualize the microphantom's test lines is the scattering strengths of the microphantoms at the two different wavelengths. \par 
Notably, both MSBP-E and MSBP-I use total-variation (TV) regularization in order to stabilize the convergence of the nonconvex iterative solver in the presence of noise \cite{strong2003edge}. Especially in the case of using \SI{633}{\nano \meter} wavelength light, TV regularization results in 3D RI reconstructions with less noise compared to the 3D reconstructions computed via GPSC, which does not use TV regularization. This can be directly visualized in the 2D cross-sections in Fig. \ref{fig:results-combined}, and is confirmed by the bounds of standard deviation shown in the 1D cross-sectional plots. However, the drawback of regularization is that it has a blurring effect on high-resolution features. Because the microphantom is less scattering in \SI{835}{\nano \meter} wavelength light, only GPSC managed to reconstruct the high spatial-frequency test lines within the microphantom. Typically, the strength of regularization is manually tuned to fit experimental factors and balance between the tradeoff between achieving iterative stability versus high imaging resolution. \par 
In terms of the average RI, the knowledge about the ground truth RI distribution of the phantom makes it possible to quantitatively compare reconstruction results with GPSC, MSBP-E, and MSBP-I. We see that all methods successfully capture the bulk characteristics of the microphantom. As described above, the 3D reconstruction via GPSC exhibits grainy artifacts when using \SI{633}{\nano \meter} wavelength light, likely due to the microphantom being multiple-scattering at that wavelength. Furthermore, MSBP-I outputs slightly overestimated RI values, and also suffers from low-frequency spatial artifacts (which have been observed in other intensity-only phase-imaging techniques \cite{wang2011spatial,tian2015quantitative}). Other works have shown that MSBP-I demonstrates higher accuracy when using partially-coherent illumination, which drastically reduces coherent noise \cite{Chowdhury2019}. Future work may include repeating this analysis across a larger range of TPM reconstruction techniques with more complex scattering microphantoms.

\section{Conclusions}
Here we presented a 3D-printed microphantom with known geometry and RI distribution, for the purpose of quantitatively assessing reconstruction quality and accuracy between various TPM reconstruction methodologies. For the microphantom presented here, the cell-like portion within the center is weakly scattering. However, this cell-like portion is surrounded with a dense pseudo-random grid of heterogenous RI structures, which scatters light. This enables the total microphantom to be an appropriate 3D calibration object to robustly evaluate the capability of various TPM methodologies in reconstructing 3D multiple-scattering samples. It should be noted that it is possible to change the scattering properties of the phantom by changing the design of the grid or by changing the refractive index of the immersion liquid. \par 
We have shown experimentally that changing the illumination wavelength affects the scattering nature of the microphantom. Specifically, though the microphantom is multiple scattering with \SI{633}{\nano \meter} wavelength light, it is weak scattering with  \SI{835}{\nano \meter} wavelength light. This naturally indicates that the optimum choice for illumination wavelength must balance between resolution ($\frac{\lambda}{NA}$ for single projection) and scattering strength. As has been shown, there are cases when instead of applying multiple-scattering methods, it is advantageous to increase the illumination wavelength (thus decreasing the scattering strength of the sample) and apply a method based on the Fourier Diffraction Theorem that does not utilize the total-variation constraint. More fundamentally, however, we showed that 3D printed microphantoms enable quantitative assessment of 3D RI reconstruction accuracy across various TPM methodologies. This capability is important when choosing a TPM method optimized for specific classes of samples and imaging conditions. An important and natural extension of this work is to develop a methodology to 3D-print complex and realistic microphantoms that mimic a wide range of biological specimens, ranging from multicellular clusters to bulk tissues and small organisms.

\section{Backmatter}

\subsection{Funding}
The research leading to the described results was carried out within the program TEAM TECH/2016-1/4 of Foundation for Polish Science, co-financed by the European Union under the European Regional Development Fund. Calibration object development was funded by FOTECH-1 project granted by Warsaw University of Technology under the program Excellence Initiative: Research University (ID-UB). We also gratefully acknowledge support from the University of Texas at Austin, Cockrell School of Engineering, and the Chan Zuckerberg Initiative.

\subsection{Acknowledgments}
The authors want to thank Demetri Psaltis and Joowon Lim from École Polytechnique Fédérale de Lausanne for providing the code for Learning Tomography algorithm.

\subsection{Disclosures}
The authors declare no conflicts of interest.

\subsection{Data availability} Data underlying the results presented in this paper are available in Dataset 1 \cite{zenodo}.

\subsection{CRediT}
\textbf{Wojciech Krauze:} Conceptualization, Formal analysis, Methodology, Software, Visualization, Writing – original draft. \textbf{Arkadiusz Kuś:} Conceptualization, Funding acquisition, Investigation, Methodology, Project administration, Visualization, Writing – original draft. \textbf{Michał Ziemczonok:} Data curation, Investigation, Methodology, Resources, Visualization, Writing – original draft. \textbf{Max Haimowitz:} Software, Writing – review \& editing. \textbf{Shwetadwip Chowdhury:} Funding acquisition, Software, Writing – review \& editing. \textbf{Małgorzata Kujawińska:} Funding acquisition, Project administration, Writing – review \& editing.

\bibliographystyle{unsrt} 
\bibliography{article}

\end{document}